\documentclass{Interspeech}

\interspeechcameraready

\title{Towards Robust Overlapping Speech Detection: A Speaker-Aware Progressive Approach Using WavLM}

\author[affiliation={1,2}]{Zhaokai}{Sun}
\author[affiliation={1}]{Li}{Zhang}
\author[affiliation={1}]{Qing}{Wang}
\author[affiliation={2}]{Pan}{Zhou}
\author[affiliation={1}{*}]{Lei}{Xie}

\affiliation{Audio, Speech and Language Processing Group (ASLP@NPU), School of Software}{Northwestern Polytechnical University}{China}
\affiliation{}{Space AI}{Li Auto}
\email{zksun@mail.nwpu.edu.cn, lxie@nwpu.edu.cn}

\usepackage{comment}
\usepackage{enumitem}
\usepackage{amsmath}
\usepackage{graphicx}

\begin{document}

\maketitle

\begin{abstract}

    Overlapping Speech Detection (OSD) aims to identify regions where multiple speakers overlap in a conversation, a critical challenge in multi-party speech processing. This work proposes a speaker-aware progressive OSD model that leverages a progressive training strategy to enhance the correlation between subtasks such as voice activity detection (VAD) and overlap detection. To improve acoustic representation, we explore the effectiveness of state-of-the-art self-supervised learning (SSL) models, including WavLM and wav2vec 2.0, while incorporating a speaker attention module to enrich features with frame-level speaker information. Experimental results show that the proposed method achieves state-of-the-art performance, with an F1 score of 82.76\% on the AMI test set, demonstrating its robustness and effectiveness in OSD.
    
\end{abstract}

\section{Introduction}

Multi-party conversation recognition is one of the most challenging problems in speech processing \cite{sell18_interspeech}, particularly due to difficulties in accurately identifying overlapping speech regions. Overlapping regions present unique challenges, as they involve both overlap of speech segments and speaker identities. Effectively addressing these challenges is critical for applications such as speaker diarization, automatic speech recognition (ASR), and multi-speaker dialogue systems \cite{Boakye08-osd, Charlet13-osdsd, diez18_interspeech}. Overlap Speech Detection (OSD) has emerged as a crucial component in tackling these challenges by identifying overlapping speech regions in complex acoustic environments.

Recent advancements in deep learning have introduced powerful tools for OSD, leveraging convolutional neural networks (CNNs), long short-term memory (LSTM) networks, and transformer architectures \cite{Kune19-10.1007/978-3-030-26061-3_26, bullock2019overlapawarediarizationresegmentationusing, wang2022spatial}. These methods process diverse input features such as raw waveforms, filter banks (FBank), spectrograms, and x-vectors \cite{bredin2021endtoendspeakersegmentationoverlapaware, yin2023largescalelearningoverlappedspeech, kazimirova2018automatic, mateju2022overlapped}. However, the robustness of these systems remains closely tied to the diversity and quality of training data. Due to the scarcity of real-world dialogue datasets with sufficient overlap durations, most OSD research relies on simulated data or selectively filtered open-source datasets. Simulated datasets, such as LibriSpeech-derived LibriHeavyMix \cite{jin2024libriheavymix}, are generated based on statistical characteristics of overlapping and non-overlapping regions, but often exhibit significant performance variability across test conditions due to domain mismatches. In \cite{kunevsova2023-wav2vecmultitask}, model trained on simulated datasets shows performance degradation on relastic test datasets. Open-source datasets, including AliMeeting \cite{yu2022m2met} and AMI \cite{Renals07-ami4430116}, also serve as benchmarks. In \cite{yin2023largescalelearningoverlappedspeech}, Yin et al. propose a large-scale benchmark, with filtered open-source datasets.

Multi-task learning (MTL) \cite{zhang2021survey} has shown promise in improving OSD performance by integrating related tasks, such as voice activity detection (VAD). Incorporating VAD as a subtask helps capture human speech patterns and enhances the overall robustness of OSD systems. However, many existing approaches treat MTL as a three-class classification problem, which often underutilizes the intrinsic relationships between tasks \cite{jung2021three}. For instance, models like CRNN \cite{zheng2021beamtransformer} classify audio frames into silence, single-speaker, and overlap categories, but this simplification can lead to imbalances in precision and recall.

In addition to temporal overlap detection, distinguishing speaker identities within overlapping regions is another critical challenge. Speaker recognition, another critical component of speech analysis, provides valuable insights for addressing overlapping speech \cite{tu2022survey}. By extracting speaker-specific features, speaker recognition systems enable applications such as access control, personalized voice services, and dialogue analysis. With the development of deep learning, the robustness of speaker recognition systems has gradually improved, as seen in systems like ecapa \cite{dawalatabad2021ecapa} and campplus \cite{wang2023cam++}. Pretrained self-supervised learning (SSL) models, such as WavLM \cite{chen2022wavlm}, have demonstrated remarkable effectiveness in capturing speaker information, particularly for overlapping audio scenarios. However, most OSD systems fail to explicitly incorporate frame-level speaker representations, missing an opportunity to improve overlap detection.

To address these challenges, we propose a robust and progressive OSD framework that leverages WavLM and incorporates speaker representations to enhance overlap detection. Our approach introduces a progressive learning strategy to better utilize task interrelationships, enabling stepwise refinement of VAD and OSD components. Specifically, we mask encoder hidden states using VAD logits and feed the masked representations into the OSD decoder. Additionally, inspired by studies highlighting the feature distribution similarities between single-speaker and overlapping segments \cite{cord2023-diarspkmse}, we design a frame-level speaker attention module. This module extracts speaker representations and integrates them with audio features from the pretrained SSL model, enriching the OSD decoder’s input.

To train our system, we utilize a combination of simulated and real-world datasets, including LibriHeavyMix, AliMeeting, and AMI. Our training strategy involves pretraining on simulated datasets followed by fine-tuning on realistic data. This two-step process, commonly used in speaker diarization, facilitates model convergence and enhances robustness across diverse scenarios \cite{Medennikov_2020-tsvad}.

The remainder of this paper is structured as follows: Section 2 details the proposed framework and OSD system configuration. Section 3 presents the experimental setup and results. Section 4 concludes the paper and discusses future directions.

\begin{figure}[bt]
  \centering
  \includegraphics[width=\linewidth]{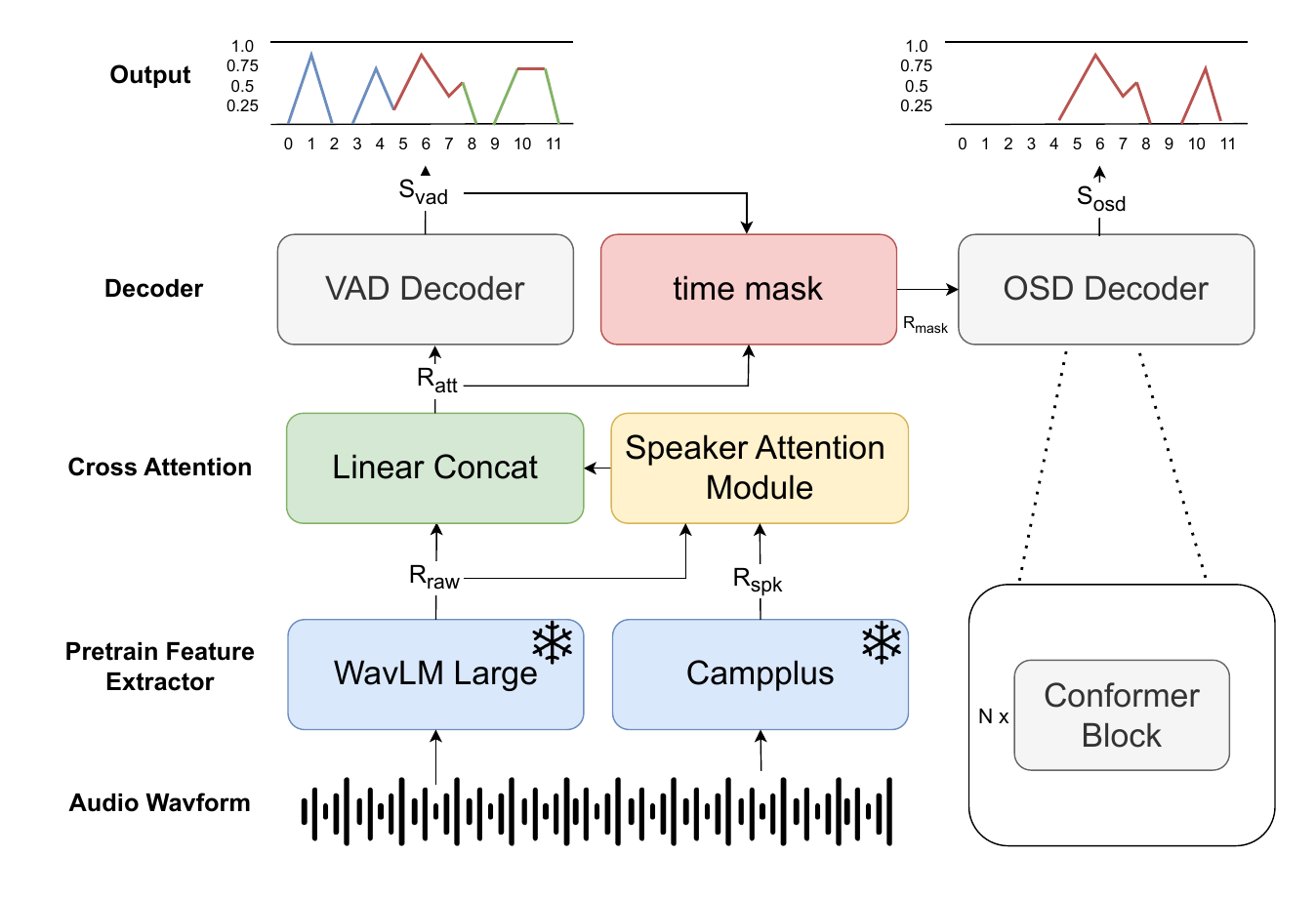}
  \caption{Our detailed progressive OSD architecture.}
  \label{fig:T-OSD model framework}
\end{figure}

\section{Methods}
In this section, we provide a comprehensive description of the proposed speaker-aware progressive OSD framework. We begin with an overview of the progressive OSD architecture, followed by a detailed explanation of the progressive OSD training strategy. Finally, we present a thorough discussion of the individual modules within the system.

\subsection{Overview}

As illustrated in Fig. \ref{fig:T-OSD model framework}, the proposed speaker-aware progressive OSD model comprises four key components: the pre-trained SSL encoder, the speaker attention module, the temporal masking module, and the decoder. During the inference process, the input audio is combined with speaker information to enrich SSL representations. These fused representations are then masked using the VAD logits before being passed through the OSD decoder.
 
In this paper, we leverage the representations generated by WavLM-Large\footnote{https://github.com/microsoft/unilm/tree/master/wavlm} to enhance OSD performance. The pre-trained WavLM model offers robust, high-level acoustic representations, as it is trained on tasks such as masked speech prediction and audio denoising \cite{chen2022wavlm}. To further refine these representations, we incorporate a speaker attention module, which enriches the WavLM hidden states with speaker-specific information. Additionally, we utilize the pre-trained CampPlus\footnote{https://www.modelscope.cn/models/iic/speech\_campplus\_sv\_zh-cn\_16k-common} model to generate frame-level speaker embeddings. These fused representations are then passed to the VAD decoder, which consists of Conformer blocks.

For the progressive learning strategy, we design the OSD input of our end-to-end system based on temporal masking, enabling the OSD decoder to focus on relevant speech segments. Specifically, the temporal mask module generates masked fusion representations by utilizing the output logits from the VAD decoder. Finally, the OSD decoder produces the OSD logits using the masked fusion representations.

\subsection{Progressive OSD modeling}

The progressive modeling approach we propose is depicted in Fig. \ref{fig:T-OSD model framework}. Previous studies typically generate VAD and OSD predictions simultaneously in the final step, applying classification losses for both tasks \cite{yin2023largescalelearningoverlappedspeech, kunevsova2023-wav2vecmultitask, lebourdais2023joint}. However, these approaches do not explicitly model the relationship between the two tasks. Serving as a speech type classification task, they share significant similarities. VAD, as a fundamental task, ensures high recall for overlapping speech detection, while the OSD model focuses on enhancing detection precision. Our goal is to improve OSD performance by explicitly modeling the relationship between VAD and OSD tasks.

In our approach, we propose a progressive strategy for OSD tasks, where the input to the OSD decoder is influenced by the VAD output. Specifically, we apply a temporal mask to the frame-level acoustic representations from the pre-trained SSL encoder using VAD logits. The temporal mask module effectively masks silent segments, enabling the OSD decoder to focus solely on speech regions, thus enhancing the model's ability to accurately distinguish overlapping speech. The time masking process is formalized in Eq \ref{eq:time mask}, where $R_{att}$ represents the hidden features enhanced by the speaker attention module, and $S_{vad}$ denotes the VAD logits.

\begin{figure}[bt]
  \centering
  \includegraphics[width=0.5\linewidth]{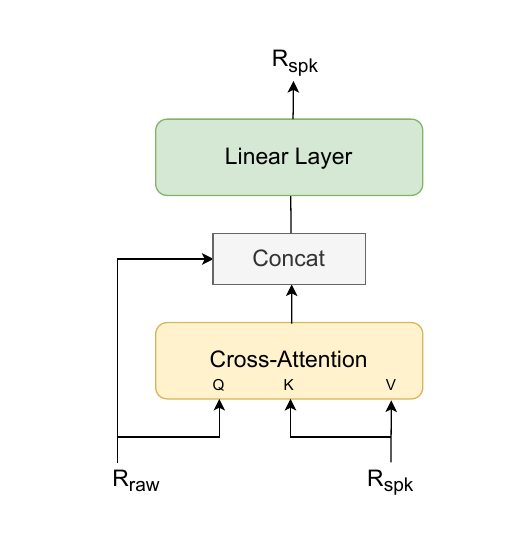}
  \caption{Frame-level speaker attention module.}
  \label{fig:T-OSD spk attention}
\end{figure}

\subsection{Pretrain SSL Encoder}
WavLM is a self-supervised speech processing model designed to leverage diverse speech datasets. It adopts an encoder-only Transformer architecture, where the encoder consists of seven convolutional layers followed by a stack of Transformer blocks with gated relative position bias \cite{chen2022wavlm}. Trained on an audio mask prediction task, WavLM has demonstrated exceptional performance across various speech processing tasks, including speaker recognition, speech recognition, and speech generation. Moreover, recent studies have explored its robustness in additional downstream applications, such as speech event detection \cite{jiang2024-wavlmspeechevent} and spoofed audio detection \cite{yang2024-wavlmdeepfakedetection}. In this paper, we adapt WavLM for OSD tasks. As illustrated in Fig. \ref{fig:T-OSD model framework}, the WavLM encoder converts input audio signals into a sequence of high-level abstract representations, serving as the foundation for subsequent processing.

The input audio waveform, denoted as $X$, with corresponding VAD, OSD labels given by $Y_{vad}$ and $Y_{osd}$. The WavLM encoder is $F_{w}$. After $T$ length audio waveform resampled to 16kHz, we extract the encoder audio representations $R_{raw}$ from pre-trained encoder as follows:
\begin{equation}
X = [x_1, x_2, ..., x_t],~x_t \in \mathbb{R}, t\in[0, T]
\label{eq:raw feature extract}
\end{equation}
\begin{equation}
Y_{vad} = [y_{vad_1}, y_{vad_2}, ..., y_{vad_t}],~y_{vad_t} \in \mathbb{R}
\label{eq:raw feature extract}
\end{equation}
\begin{equation}
Y_{osd} = [y_{osd_1}, y_{osd_2}, ..., y_{osd_t}],~y_{osd_t} \in \mathbb{R}
\label{eq:raw feature extract}
\end{equation}
\begin{equation}
R_{raw} = F_w(X)
\label{eq:raw feature extract}
\end{equation}

\subsection{Speaker Attention Module}
Previous research on overlapping speech in diarization tasks has shown that speaker representations of overlap frames exhibit correlations between overlapping speakers \cite{cord2023-diarspkmse}. Motivated by this, we incorporate speaker representations into our system to enhance OSD performance. Specifically, we utilize CampPlus, a pretrained speaker verification (SV) model based on the D-TDNN architecture, which has achieved state-of-the-art performance on SV benchmark datasets \cite{wang2023cam++}. Since WavLM is not explicitly optimized for SV tasks, we introduce a speaker attention module to refine WavLM representations by integrating frame-level speaker characteristics extracted from CampPlus.

The Fbank process is denoted as $F_{mel}$. Given an input audio waveform $X$,  the computation of the Fbank features $X_{fbank}$ follows the procedure described in \ref{eq:fbank}. The CampPlus model, $F_{sv}$, transforms the Fbank features $X_{fbank}$ into frame-level speaker representations $R_{spk}$. Subsequently, the cross-attention module $F_{att}$ integrates these speaker representations to enhance the pretrained features $R_{raw}$. Within the cross-attention module, $R_{raw}$ serves as the query vector, while $R_{spk}$ is used as both the key and value vector..

\begin{equation}
X_{fbank} = F_{mel}(X)
\label{eq:fbank}
\end{equation}
\begin{equation}
R_{spk} = F_{sv}(X_{fbank})
\label{eq:raw feature extract}
\end{equation}
\begin{equation}
R_{att} = F_{att}(R_{raw}, R_{spk}, R_{spk})+R_{raw}
\label{eq:raw feature extract}
\end{equation}

\subsection{Temporal Masking Module}
VAD and OSD tasks exhibit a hierarchical relationship, where OSD can leverage the single-speaker segments identified by VAD to focus on learning overlapping speech features. Unlike conventional multi-task learning, which produces a unified three-class prediction at the backend, our method explicitly models the interaction between VAD and OSD through a temporal masking module. In this module, a mask is applied to the fused audio representations obtained in earlier stages, based on the logits generated by the VAD module. The OSD decoder then computes the OSD scores for each frame using these masked representations.

The VAD logits are donted as $S_{vad}$. The temporal masking module, $F_{mask}$ , takes the fused audio representation $R_{att}$ along with the VAD logits to produce the masked representation $R_{mask}$. This process is formally defined by the following equation:
\begin{equation}
R_{mask} = F_{mask}(S_{vad},~ R_{att})
\label{eq:time mask}
\end{equation}

\subsection{Decoder}
The decoder module transforms the frame-level features extracted in previous stages into prediction scores. In OSD tasks, the decoder is typically composed of multiple linear layers, LSTM networks, or Transformer-based architectures. In our system, we implement the decoder using several Conformer blocks. Given that the OSD task is trained using a progressive learning strategy, separate decoders are employed for VAD and OSD. Both decoders share the same network architecture within our framework.

The VAD decoder $F_{vad}$ utilizes the attention-enhanced representation $R_{att}$ to produce VAD logits $S_{vad}$ whereas the OSD decoder $F_{osd}$ operates on the masked representation $R_{mask}$ to generate OSD logits $S_{osd}$. 
\begin{equation}
S_{vad} = F_{vad}(R_{att})
\label{eq:raw feature extract}
\end{equation}
\begin{equation}
S_{osd} = F_{osd}(R_{mask})
\label{eq:raw feature extract}
\end{equation}

To address the uncertainty introduced by human annotations, we employ a fuzzy labeling strategy. In this approach, overlapping speech frames are assigned a label of 1, with a linear decay applied near the boundaries to reflect the gradual transition between speech regions. Given this label design, our two-stage model is trained as a regression task using mean squared error (MSE) loss. Prior research \cite{kunevsova2023-wav2vecmultitask} has demonstrated the effectiveness of fuzzy labeling in OSD tasks. The frame-level labels are denoted as $y$, and the loss functions for both tasks are formulated as follows.

\begin{equation}
L_{vad} = \frac{1}{T} \sum_{i=1}^{T} ||y_{vad_i} - S_{vad_i}||,~S_{vad_i} \in S_{vad}
\label{eq:raw feature extract}
\end{equation}
\begin{equation}
L_{osd} = \frac{1}{T} \sum_{i=1}^{T} ||y_{osd_i} - S_{osd_i}||,~S_{osd_i} \in S_{osd}
\label{eq:raw feature extract}
\end{equation}

        
    

\section{Experiments}
\subsection{Experiments setup}
In this section, we describe the experimental setup used for our paper, including the datasets and training configurations.
\subsubsection{Data Corpus}

An ideal OSD system should be robust to variations in language, speaking style, and speaker-to-microphone distance. To enhance generalization, we compile a diverse multi-speaker conversational dataset from various open-source sources. During training, the model is first pretrained for five epochs on LibriHeavyMix \cite{jin2024libriheavymix} before being fine-tuned using realistic data. The details of the datasets are presented in Table \ref{tab:T-OSD datasets}.

One major challenge in OSD training is the inherent class imbalance, as overlapping speech occurs far less frequently than non-overlapping speech. To mitigate this issue, we manually curate the training data to ensure a balanced distribution, maintaining a 1:1:1 ratio among silent segments, single-speaker segments, and overlapping speech segments within each session.

Furthermore, to account for annotation uncertainty, we apply a fuzzy labeling strategy near VAD and OSD boundaries. Specifically, the overlapping speech labels are gradually decayed from 1 to 0 over a span of 10 frames near the boundary, facilitating a smoother transition and improving model robustness.

\subsubsection{Training Configuration}
The speech data is segmented into 5-second intervals. To ensure compatibility with the pretrained frontend, the audio frame length is set to 25 ms with a frame shift of 20 ms. All experimental setups maintain consistent batch sizes and optimizer configurations. Specifically, we employ the Adam optimizer with an initial learning rate of 1e-4 for training all OSD models. To prevent the model from converging to suboptimal solutions due to a fixed learning rate, we apply a weight decay of 1e-4 and utilize a cosine scheduler to dynamically adjust the learning rate, facilitating a smoother optimization process.

Additionally, we investigate the impact of reverberation augmentation during training. However, our experiments indicate that adding reverberation does not yield significant performance improvements; therefore, it is not incorporated into our final training pipeline.

\subsection{Experiment Results}

\begin{table}[tb]
    \centering
    \caption{\textnormal{Details of datasets. *LibriheavyMix~\cite{jin2024libriheavymix} is an open-source simulation conversastion dataset.}}
    \begin{tabular}{l l l l l}
        \toprule
        \textbf{Dataset} & \textbf{Style}  & \textbf{Hours} & \textbf{Overlap Ratio~(\%)} \\
        \midrule
        Alimeeting & Meeting/Far  & 104.75 & 42.27\\
        AMI & Meeting/Far  & 75 & 19 \\
        LibriheavyMix\textsuperscript{*} & Multi/Near & 240 & 42.56\\
        \bottomrule
    \end{tabular}
    \label{tab:T-OSD datasets}
\end{table}

\begin{table}[tb]
    \centering
    \caption{\textnormal{Results(\%) of multi-task model on AMI test. }}
    \begin{tabular}{l ccc ccc}
        \toprule
        
        \textbf{Method} & \textbf{Recall} & \textbf{Precision} & \textbf{F1}\\
        \midrule
        CNN~[2] & 44.6 & 75.8 & 56.1\\  
        x-vectors~[15] & 47.7 & 85.5 & 61.0\\  
        pyannote~[16] & 80.7 & 70.5 & 75.3\\  
        conformer~[17] & 65.03 & 64.43 & 64.73\\  
        xlsr-conformer~[18] & 79.38 & 79.04 & 79.21\\  
        \midrule
        \textbf{ours~(progressive)} & \textbf{81.48} & \textbf{84.08} & \textbf{82.76}\\ 
        \bottomrule
    \end{tabular}
    
    \label{tab:competitive results}
\end{table}

\begin{table}[tb]
    \centering
    \caption{\textnormal{Results(\%) of speaker ablation experiments. }}
    \begin{tabular}{l ccc ccc}
        \toprule
        
        \textbf{Method} & \textbf{Recall} & \textbf{Precision} & \textbf{F1}\\
        \midrule
        \textbf{p-OSD-wavlm-spkAtt(ours)} & \textbf{81.48} & \textbf{84.08} & \textbf{82.76}\\ 
        \midrule
        p-OSD-wavlm-spkMSE & 80.80 & 82.45 & 81.62\\ 
        p-OSD-wavlm & 79.43 & 79.52 & 79.47\\    
        \bottomrule
    \end{tabular}
    
    \label{tab:spk ablation}
\end{table}

\begin{table}[tb]
    \centering
    \caption{\textnormal{Results(\%) of progressive ablation experiments. e.g. p/u means progressive or unified approach.}}
    \begin{tabular}{l ccc ccc}
        \toprule
        
        \textbf{Method} & \textbf{Recall} & \textbf{Precision} & \textbf{F1}\\
        \midrule
        \textbf{p-OSD-wavlm-spkAtt~(ours)} & \textbf{81.48} & \textbf{84.08} & \textbf{82.76}\\ 
        u-OSD-wavlm-spkAtt & 80.90 & 83.55 & 82.20\\ 
        \midrule
        p-OSD-wavlm-spkMSE & 80.80 & 82.45 & 81.62\\ 
        u-OSD-wavlm-spkMSE & 80.52 & 81.92 & 81.21\\
        \midrule
        p-OSD & 66.82 & 65.10 & 65.95\\
        u-OSD & 65.03 & 64.43 & 64.73\\
        \bottomrule
    \end{tabular}
    
    \label{tab:progressive ablation}
\end{table}

\begin{table}[tb]
    \centering
    \caption{\textnormal{Results(\%) of SSL encoder ablation experiments.}}
    \begin{tabular}{l ccc ccc}
        \toprule  
        \textbf{Method} & \textbf{Recall} & \textbf{Precision} & \textbf{F1}\\
        \midrule
        \textbf{p-OSD-wavlm-spkAtt~(ours)} & \textbf{81.48} & \textbf{84.08} & \textbf{82.76}\\ 
        p-OSD-xlsr-spkAtt & 79.19 & 80.97 & 80.07\\ 
        \midrule
        p-OSD-wavlm & 79.43 & 79.52 & 79.47\\    
        p-OSD-xlsr & 78.45 & 79.03 & 78.73\\
        \midrule
         P-OSD & 66.82 & 65.10 & 65.95 \\
        \bottomrule
    \end{tabular}    
    \label{tab:ssl ablation}
\end{table}

To comprehensively investigate the effectiveness of our approach, we conduct four main experiments. Table \ref{tab:competitive results} provides a comparison between our system and the most competitive existing methods. Table \ref{tab:spk ablation} examines the impact of the speaker attention module. Table \ref{tab:progressive ablation} evaluates the improvements introduced by the progressive learning strategy across different system configurations. Finally, Table \ref{tab:ssl ablation} explores the effect of different pretrained SSL encoders on feature extraction by substituting alternative models.


The first experiment aims to evaluate the performance of the proposed progressive OSD system in comparison with related studies. As shown in Table \ref{tab:competitive results}, our system achieves an F1 score of 82.76\%, outperforming all previously reported methods. Notably, it demonstrates a relative improvement of 4.4\% over the previous state-of-the-art system, XLSR-Conformer. Additionally, both recall and precision scores show significant enhancements, indicating that the progressive learning approach effectively boosts overall system performance.

The second experiment examines different methods for integrating speaker information into the OSD system. In our proposed framework, speaker information is incorporated through the speaker attention module. To provide a comparative analysis, we also implemented a previously proposed approach that aligns OSD representations with speaker features using mean squared error (MSE) loss \cite{cord2023-diarspkmse}. As presented in Table \ref{tab:spk ablation}, the progressive OSD system with the speaker attention module achieves an F1 score of 82.76\% on the AMI test set, while the system using MSE attains 81.62\%. These findings suggest that, compared to MSE-based alignment, the attention mechanism enables more effective utilization of speaker information, leading to improved system performance.

The third experiment aims to demonstrate the effectiveness of our two-stage approach in improving recall and reducing the false alarm rate in the OSD task. To ensure a fair comparison, we modified the multi-task modeling strategy used in previous OSD systems into a progressive approach while maintaining an identical modeling framework. As shown in Table \ref{tab:progressive ablation}, our progressive OSD system, incorporating WavLM and the speaker attention module, achieves the highest F1 score, with a relative improvement of 0.6\%. The comparison between the unified and progressive approaches indicates that the progressive model yields superior recall and false alarm rates, further validating the effectiveness of the proposed progressive learning strategy.

\section{Conclusion}

In this paper, we propose a speaker-aware progressive OSD model that leverages a progressive learning strategy to effectively exploit the correlations between subtasks, thereby enhancing the robustness and performance of OSD systems. Experimental results consistently demonstrate the advantages of this approach across different configurations. By integrating a pretrained SSL frontend with a speaker attention module, the system achieves enriched acoustic representations, further boosting detection accuracy. The proposed speaker-aware progressive OSD model achieves superior performance compared to previous state-of-the-art methods, underscoring the effectiveness of the progressive learning strategy and the speaker-enhanced acoustic feature design.

\ifinterspeechfinal
     The Interspeech 2025 organisers
\else
    
\fi

\bibliographystyle{IEEEtran}
\bibliography{template}

\end{document}